\def\mode{1}
\if 0\mode
  \documentclass[english,titlepage,oneside,12pt]{article}
\else
  \documentclass[conference]{IEEEtran}
\fi

\usepackage[T1]{fontenc}
\pdfoutput=1
\usepackage{graphicx}
\usepackage{setspace}
\usepackage{subcaption}

\usepackage{color}
\usepackage{multirow}
\usepackage{multicol}

\usepackage{booktabs} 
\usepackage{tabularx}

\usepackage{algpseudocode}
\usepackage{algorithm}
\usepackage{float}
\newfloat{algorithm}{t}{lop}

\usepackage{eqparbox}
\usepackage{gensymb}
\usepackage{amsmath}
\usepackage{amssymb}
\usepackage{amsfonts}

\usepackage{url}
\usepackage[capitalise,noabbrev,nameinlink]{cleveref}

\graphicspath{ {./} }

\makeatletter

\if 0\mode

\usepackage[tiny,rm]{titlesec}
\newpagestyle{trbstyle}{
\sethead{Wilbur, Ayman, Ouyang, Poon, Kabir, Vadali, Pugliese, Freudberg, Laszka and Dubey}{}{\thepage}
}
\renewcommand*{\refname}{\uppercase{References}}
\titleformat{\section}{\bfseries}{}{0pt}{\uppercase}
\titlespacing*{\section}{0pt}{12pt}{*0}
\titleformat{\subsection}{\bfseries}{}{0pt}{}
\titlespacing*{\subsection}{0pt}{12pt}{*0}
\titleformat{\subsubsection}{\itshape}{}{0pt}{}
\titlespacing*{\subsubsection}{0pt}{12pt}{*0}

\usepackage{enumitem}
\setlist[1]{labelindent=0.5in,leftmargin=*}
\setlist[2]{labelindent=0in,leftmargin=*}

\else
\fi

\usepackage{amsmath}
\renewcommand{\fnum@figure}{\textbf{FIGURE~\thefigure} }
\renewcommand{\fnum@table}{\textbf{TABLE~\thetable} }

\usepackage[sort,numbers]{natbib}

\usepackage{calculator}
\usepackage{totcount}
\regtotcounter{table} 	
\regtotcounter{figure} 	

\usepackage{newtxtext,newtxmath}

\usepackage[T1]{fontenc}
\usepackage{textcomp}

\usepackage{graphicx}
\usepackage[pagewise]{lineno}

\date{} 

\makeatother

\usepackage[utf8]{inputenc}
\usepackage{pgfplots}
\usepackage{pgfplotstable}
\usetikzlibrary{pgfplots.dateplot}
\usetikzlibrary{pgfplots.statistics}
\pgfplotsset{compat=1.14}

\pgfplotsset{
compat=1.14,
SmallBarPlot/.style={
    font=\footnotesize,
    ybar,
    width=\linewidth,
    ymin=0,
    xtick=data,
    xticklabel style={text width=0.8cm, align=center},
    xtick pos=left
},
BlueBigBars/.style={
    fill=blue!40, bar width=0.25
},
RedBigBars/.style={
    fill=red!40, bar width=0.25
},
BlueBars/.style={
    fill=blue!40, bar width=0.1
},
RedBars/.style={
    fill=red!40, bar width=0.1
},
SmallBlueBars/.style={
    fill=blue!40, bar width=0.05
},
SmallYellowBars/.style={
    fill=yellow!40, bar width=0.05
},
SmallRedBars/.style={
    fill=red!40, bar width=0.05
},
SmallGreenBars/.style={
    fill=green!40, bar width=0.05
},
SmallBlackBars/.style={
    fill=black!40, bar width=0.05
},
}

\usepackage[textsize=tiny,textwidth=1.54cm]{todonotes}

\clearpage
\usepackage[english]{babel} 
\begin{document}
\if 0\mode

\thispagestyle{empty}
\title{\vspace{-3.5cm}Impact of COVID-19 on Public Transit\\Accessibility and Ridership}
\author{
{\large Michael Wilbur (corresponding author)}\vspace{-1ex}\\
{\normalsize Department of Electrical Engineering and Computer Science}\vspace{-1ex}\\
{\normalsize Institute for Software Integrated Systems}\vspace{-1ex}\\
{\normalsize Vanderbilt University }\vspace{-1ex}\\
{\normalsize michael.p.wilbur@vanderbilt.edu }\\
{\normalsize{} }\\
{\large Afiya Ayman}\vspace{-1ex}\\
{\normalsize Department of Computer Science}\vspace{-1ex}\\
{\normalsize University of Houston }\vspace{-1ex}\\
{\normalsize{} }\\
{\large Anna Ouyang}\vspace{-1ex}\\
{\normalsize Department of Electrical Engineering and Computer Science}\vspace{-1ex}\\
{\normalsize Vanderbilt University }\vspace{-1ex}\\
{\normalsize{} }\\
{\large Vincent Poon}\vspace{-1ex}\\
{\normalsize Department of Computer Science}\vspace{-1ex}\\
{\normalsize University of Houston }\vspace{-1ex}\\
{\normalsize{} }\\
{\large Riyan Kabir}\vspace{-1ex}\\
{\normalsize Department of Electrical Engineering and Computer Science}\vspace{-1ex}\\
{\normalsize Vanderbilt University }\vspace{-1ex}\\
{\normalsize{} }\\
{\large Abhiram Vadali}\vspace{-1ex}\\
{\normalsize Department of Electrical Engineering and Computer Science}\vspace{-1ex}\\
{\normalsize Vanderbilt University }\vspace{-1ex}\\
{\normalsize{} }\\
{\large Philip Pugliese}\vspace{-1ex}\\
{\normalsize Chattanooga Area Regional Transportation Authority}\vspace{-1ex}\\
{\normalsize{} }\\
{\large Daniel Freudberg}\vspace{-1ex}\\
{\normalsize Nashville Metropolitan Transit Authority}\vspace{-1ex}\\
{\normalsize{} }\\
{\large Aron Laszka}\vspace{-1ex}\\
{\normalsize Department of Computer Science}\vspace{-1ex}\\
{\normalsize University of Houston }\vspace{-1ex}\\
{\normalsize{} }\\
{\large Abhishek Dubey}\vspace{-1ex}\\
{\normalsize Department of Electrical Engineering and Computer Science}\vspace{-1ex}\\
{\normalsize Institute for Software Integrated Systems}\vspace{-1ex}\\
{\normalsize Vanderbilt University }\vspace{-1ex}\\
{\normalsize{} }\\
}
\thispagestyle{empty}

\maketitle

\newpage
\thispagestyle{empty}

\else
\title{Impact of COVID-19 on Public Transit\\Accessibility and Ridership}
\author{\IEEEauthorblockN{Michael Wilbur\IEEEauthorrefmark{1},  Afiya Ayman\IEEEauthorrefmark{2}, Anna Ouyang\IEEEauthorrefmark{1}, Vincent Poon\IEEEauthorrefmark{2}, Riyan Kabir\IEEEauthorrefmark{1}, Abhiram Vadali\IEEEauthorrefmark{1}, \\Philip Pugliese\IEEEauthorrefmark{3}, Daniel Freudberg\IEEEauthorrefmark{4}, Aron Laszka\IEEEauthorrefmark{2}, Abhishek Dubey\IEEEauthorrefmark{1}}
\IEEEauthorblockA{\IEEEauthorrefmark{1}\textit{Vanderbilt University}, \IEEEauthorrefmark{2}\textit{University of Houston}, \IEEEauthorrefmark{3}\textit{Chattanooga Area Regional Transportation Authority}, \\\IEEEauthorrefmark{3}\textit{Nashville Metropolitan Transit Authority}}
}

\maketitle
\fi

\section*{Abstract}

Public transit is central to cultivating equitable communities. Meanwhile, the novel coronavirus disease COVID-19 and associated social restrictions has radically transformed ridership behavior in urban areas. Perhaps the most concerning aspect of the COVID-19 pandemic is that low-income and historically marginalized groups are not only the most susceptible to economic shifts but are also most reliant on public transportation. As revenue decreases, transit agencies are tasked with providing adequate public transportation services in an increasingly hostile economic environment. Transit agencies therefore have two primary concerns. First, how has COVID-19 impacted ridership and what is the new post-COVID normal? Second, how has ridership varied spatio-temporally and between socio-economic groups? In this work we provide a data-driven analysis of COVID-19's affect on public transit operations and identify temporal variation in ridership change. We then combine spatial distributions of ridership decline with local economic data to identify variation between socio-economic groups. We find that in Nashville and Chattanooga, TN, fixed-line bus ridership dropped by 66.9\% and 65.1\% from 2019 baselines before stabilizing at 48.4\% and 42.8\% declines respectively. The largest declines were during morning and evening commute time. Additionally, there was a significant difference in ridership decline between the highest-income areas and lowest-income areas (77\% vs 58\%) in Nashville. 

\hfill\break
\noindent\textbf{Keywords}: COVID-19, ridership, socio-economics, spatio-temporal

\if 0\mode
  \newpage
\else
  
\fi

\section{Introduction}
\label{sec:intro}

The novel coronavirus COVID-19 and the associated social restrictions have radically transformed travel behavior in urban areas throughout the world. While COVID-19 has transformed normal operations in almost all industries, the social distancing measures and precautions associated with this virus have had particularly devastating effects on public transit. For instance, since the World Health Organization (WHO) declared COVID-19 a pandemic on March 11, 2020 \cite{whopandemicannouncement} subway ridership in New York City has dropped by upwards of 91\% \cite{a2c2smartwhitepaper2}. This disruption has created pressing operational challenges for public transit agencies. 

Foremost, agencies must determine how to continue providing adequate service while navigating a rapidly changing environment with reduced resources. This involves first quantifying the affects of the pandemic to date. However, the decentralized nature of government policies and recommendations in the United States makes it challenging to identify global solutions for local transit agencies. Therefore local agencies must identify solutions tailored to their unique circumstances. Additionally, the local outlook is dynamic as regulations are lifted or restricted over time. Therefore careful data-driven modeling and analysis is required to stay up-to-date on local ridership behavior as well as the financial effects going forward.

Perhaps the most concerning aspect of the COVID-19 pandemic is that low-income and historically marginalized groups are not only the most likely to be affected financially by the crisis but are also the least likely to own their own cars \cite{a12brown2017car}. Therefore most rely on the public transit system to get to work, school or access child services. Additionally, as most people working in grocery stores, logistics and cleaning have been labeled ``essential services,'' many people from low-income groups do not have the luxury of working remotely from home. As resources are limited by drastic drops in ridership, agencies must take care in identifying trips to be cut so as to not hurt those most reliant on local transit services. 

This work is primarily concerned with two questions. First, to what degree has the COVID-19 pandemic and associated state restrictions affected ridership and operations of fixed-line public transit? We focus on Nashville and Chattanooga, TN. We present total lost riders over time compared to 2019 baselines and provide a spatio-temporal analysis of ridership decline throughout both cities. Secondly, are there disparities in ridership changes across socio-economic groups? For this we provide a spatial analysis of ridership patterns throughout the COVID-19 pandemic and correlate our findings with publicly available economic data to draw conclusions regarding changes in behavior between socio-economic groups.

\section{Contributions and Key Findings}
The primary contributions of this work are as follows:
\begin{enumerate}
    \item We provide a summary of ridership changes due to COVID-19 in Nashville and Chattanooga, TN. We find that ridership dropped by up to 66.9\% and 65.1\% in Nashville and Chattanooga respectively by late April before starting a moderate recovery.
    \item We performed a temporal investigation of ridership pre and post-COVID-19 and find that an out-sized proportion of changes in ridership occur during weekdays during morning and evening rush, indicating that Stay at Home orders and remote work options are a significant factor in ridership declines.
    \item Our spatial analysis indicates that change in ridership varies greatly between census tracts and neighborhoods. By incorporating economic data at the census tract level we found that ridership declined up to 19\% more in high-income neighborhoods than in the lowest income parts of Nashville.
\end{enumerate}

The remainder of this article is as follows. First, we summarize recent literature regarding the impact of COVID-19 on public transit systems and socio-economic transportation studies in \cref{sec:relatedWork}. Then we describe the data and processing methods in \cref{sec:dataCollection}. In \cref{sec:results}, we outline our analysis methods and results. We address possible limitations of this work in \cref{sec:threats} and 
finally we summarize our findings and discuss future work in \cref{sec:conclusion}.


\section{Related Work}
\label{sec:relatedWork}
In this section we cover literature related to COVID-19 in the context of transportation systems and the interaction of socio-economics on transit usage.

\subsection{COVID-19 and transportation}

Fixed-line bus and rail public transit inherently involves moving passengers in an enclosed space. One of the major reasons there has been such significant declines in public transit ridership is the fear of COVID-19. In public health fields, the study of infectious disease transmission through public transit and air travel is well studied \cite{browne2016roles}, \cite{andrews2013modeling}, \cite{zhao2013transportation}, \cite{bota2017modeling}. While there is a growing number of publications regarding the spread of COVID-19 by air travel \cite{chinazzi2020effect}, there is a lack of information on how this applies to public transit \cite{atlantic}. Regardless of transmission rates on public transit ridership has declined significantly as we show in this work.

Given how fast COVID-19 has transformed life in urban areas the limited amount of work related to virus in the context of public transit ridership mostly consists of pre-print publications and government public releases. The Connected Cities With Smart Transportation (C2SMART) group at New York University has released monthly whitepapers related to the impacts of COVID-19 on New York City and Seattle. They find that in New York City, average subway and commuter rail ridership is down 80\% while bus ridership is down 50\% in the first week of July, 2020 with a peak subway ridership decline of 94\% in late March \cite{a1c2smartwhitepaper1,a2c2smartwhitepaper2,a3c2smartwhitepaper3,a4c2smartwhitepaper1}. There are similar findings in European cities \cite{aloi2020effects}.

There has been some recent work investigating mode shift away from public transit. While modeling lasting effects of the pandemic is in its early stages, in some high transit cities even moderate shifts from public transit to personal vehicles can increase travel times by 5 to 10 minutes on average for one way trips \cite{a5hu2020impacts}. On the other hand, in New York City the bike sharing program CitiBike has been more resilient to loss in ridership than the subway system and there is some evidence of transit users shifting to the shared bike programs \cite{teixeira2020link}.

\subsection{Socio-economics in transportation}
Previous research indicates different transit behaviors among socioeconomic classes. When it comes to public transit, low-income and historically marginalized groups are particularly reliant on public transportation. Minorities and low-income households account for 63\% of transit riders in the United States \cite{pucher2003socioeconomics}. Additionally, low-income groups are more likely to ride buses while high income individuals are more likely to utilize rail systems \cite{a10grahn2019socioeconomic}. According to a 2017 publication from the American Public Transportation Association, 30\% of bus riders have a household income of less than \$15,000, while 12\% of bus riders have a household income of \$100,000 or more. Among rail riders, only 13\% have household incomes below \$15,000, while 29\% have household incomes of \$100,000 or more \cite{a9apta}. In terms of public versus private transit, a study conducted in Hawaii reported key differences between bus riders and solo drivers. The mean household income of a bus rider was 16\% lower than that of a solo driver \cite{a11flannelly1989multivariate}. Bus riders also, on average, owned fewer cars per household (1.7 cars) compared to solo drivers (2.3 cars) \cite{a11flannelly1989multivariate}. 

A major reason low-income groups are heavily reliant on public transportation is their likelihood of owning a personal vehicle. According to an analysis of 2012 California Household Travel Survey data, 78\% of households without a car do not have a car as a result of economic or physical barriers \cite{a12brown2017car}. Together, these studies suggest that individuals of a lower socioeconomic background may be disproportionately impacted by changes in public transit availability. It is important to note that these trends are not unique to the United States; a case study conducted in France found that low income individuals comprised a larger portion of public transit ridership than high income individuals \cite{a13mohamed2014understanding}.

\section{Data Collection and Processing}
\label{sec:dataCollection}

In this section we outline the datasets used in this work which consist of ridership boarding information, economic data per census tract and COVID-19 cases per day as well as data processing and filtering.

\subsection{Ridership data}

\if 0\mode
\begin{table}[]
\caption{Boarding counts before and after processing and number of census tracts for Nashville and Chattanooga datasets.}
\centering
\begin{tabular}{|c|c|c|c|}
\hline
            & \begin{tabular}[c]{@{}c@{}}\textbf{Raw Boardings} \\ \textbf{(2020 YTD)}\end{tabular} & \begin{tabular}[c]{@{}c@{}}\textbf{Processed Boardings} \\ \textbf{(2020 YTD)}\end{tabular} & \textbf{Number of Census Tracts} \\ \hline
\textbf{Nashville}   & 2,800,000   & 2,800,000   & 161        \\ \hline
\textbf{Chattanooga} & 464,570     & 445,987   & 82            \\ \hline
\end{tabular}
\label{table:processing}
\end{table}
\else
\begin{table}[]
\caption{Boarding counts before and after processing and number of census tracts for Nashville and Chattanooga datasets.}
\centering
\begin{tabular}{|c|c|c|}
\hline
Metric                                                                    & Nashville & Chattanooga \\ \hline
\begin{tabular}[c]{@{}c@{}}Raw Boardings \\ (2020 YTD)\end{tabular}       & 2,800,000 & 464,570     \\ \hline
\begin{tabular}[c]{@{}c@{}}Processed Boardings \\ (2020 YTD)\end{tabular} & 2,800,000 & 445,987     \\ \hline
\# of Census Tracts                                                       & 161       & 82          \\ \hline
\end{tabular}
\label{table:processing}
\end{table}
\fi


Boarding count data was provided by the Metropolitan Government of Nashville and Davidson Count for the fixed-line bus systems of Nashville from January 1, 2019 to July 1, 2020. Boarding information was also acquired from the Chattanooga Area Regional Transportation Agency (CARTA) between January 1, 2020 to July 1, 2020. The ridership data was derived from farebox units on all passenger vehicles servicing trips within these time ranges. Farebox included a record of reach passenger boarding event. It also included driver information, shift changes and when vehicles switch routes. This information was filtered so that only boarding events remained. In 2020 there were 2.8 million documented boardings in Nashville between January 1, 2020 and July 1, 2020 and for Chattanooga there were 465k documented boardings between January 1, 2020 and July 1, 2020. Each row in the respective datasets corresponded to a single boarding event.

As complete data was available for Nashville, TN in 2019 we derived baseline ridership metrics by comparing weekly data in 2020 directly to the corresponding week in 2019. Additionally, the full 2019 data provided GPS locations which allowed for spatial comparisons to baseline ridership. For Chattanooga we were provided with aggregated monthly total boardings in 2019. For baseline calculations we compared each week with the average ridership per week in that month from 2019. 

For Nashville, the GPS location of the vehicle at the time of boarding was available for each boarding event. However for Chattanooga, missing GPS readings were significant. Therefore to add GPS locations to the ridership data in Chattanooga we joined the ridership data with a separate telemetry dataset from on-board devices provided by ViriCiti \cite{viriciti}. For each boarding event we used the unique vehicle identifier to find the nearest GPS reading in the ViriCiti dataset. We filtered out boarding events that did not have a GPS reading within a 60 second window of the boarding event. After this process we found that approximately 4\% of ridership boardings were removed from the Chattanooga ridership dataset. Once the ridership datasets were prepared, we used the GPS location of each boarding event to assign that event to a 2010 Census Tract for the spatial analysis provided in \cref{subsec:spatial} and \cref{subsec:socio}.

\subsection{Economic data and COVID-19 new case counts}
Economic data was retrieved from the United States Census Bureau \cite{censusbureau} and ProximityOne \cite{proximityone}. For each 2010 census tract these sources provided a breakdown of racial demographics, income levels and housing information. New COVID-19 cases per day for Nashville and Chattanooga were retrieved from The New York Times COVID-19 Dashboard \cite{covidcountsnewyorktimes} between January 1, 2020 and July 1, 2020.

\subsection{Mapping boarding events to census tracts}
To incorporate the census tract level economic data, each boarding event was mapped to the corresponding census tract where that boarding occurred. As each census tract included a geometric polygon representing the tract this was a simple spatial join. One limitation of working with aggregated 2019 data for Chattanooga was that we could not get baseline ridership information at the census tract level. For Nashville baseline 2019 ridership at the census level was available.

\section{Analysis and Results}
\label{sec:results}

In this section we outline the main analysis and results for this work. We start by giving a high level overview of COVID-19 restrictions and the corresponding operational changes implemented by the transit agencies in Nashville and Chattanooga before moving into our analysis of ridership declines in both cities.

\subsection{COVID-19 restrictions and operational changes}

Nashville and Chattanooga both receive guidance regarding COVID-19 related restrictions directly from the State of Tennessee and also are available to impose their own regulations in excess of the state's recommendations. On March 5th the first COVID-19 case was identified in Tennessee and on March 8th the first COVDI-19 case was found in Nashville. The State of Tennessee ordered a State of Emergency regarding the pandemic on March 12, 2020 and a Safer at Home order on March 30, 2020 which mandated residents of the state stay in their homes other than for "essential activities". The Tennessee Safer at Home order ended on April 30 \cite{tennhealth}.

Nashville regulations were more swift. Nashville imposed their own Stay at Home order on March 22 which was not lifted until Phase 1 reopening began on May 11 which included allowing gatherings of up to 10 people while most businesses were allowed to open at 50\% capacity. On May 25 Nashville moved to Phase 2 which allowed gatherings of up to 25 people and most businesses could operate at 75\% capacity \cite{nashmetrohealth}. Nashville moved to a Phase 3 opening on June 20, 2020 which included provisions a limited opening for small venues (up to 250 people) however reverted back to a Phase 2 opening on July 3, 2020. 


Both Nashville and Chattanooga reduced the total number of trips in reaction to COVID-19. Unique trip identifiers were not available in either dataset. Therefore to tally the number of trips serviced per week we grouped the data by date, unique driver ID, unique vehicle ID, route and direction. Chattanooga moved to a reduced bus schedule in the middle of April while Nashville switched to a reduced schedule on March 29, 2020. Prior to the schedule change, Chattanooga serviced 781 weekly fixed-line bus trips. During the week of April 19th Chattanooga switched all weekdays to their Saturday schedule. Between April 19th and July 1st, Chattanooga serviced 373 trips per week. Nashville switched to a reduced schedule during the week of April 5. Prior to switching Nashville serviced 1954 weekly trips which was reduced to 1035 weekly trips. As we see in \cref{subsec:impacts of covid}, the most significant drops in ridership occurred well before either city moved to their respective reduced schedules.


\if 0\mode
  \subsection{Impact of COVID-19 on city-wide ridership}\label{subsec:impacts of covid}
  \if 0\mode

\begin{figure*}[]
    \centering
    \begin{subfigure}{\linewidth}
        \centering
\pgfplotsset{scaled y ticks=false}
\begin{tikzpicture}
\begin{axis}[
    width=.8\linewidth,
    height=.3\linewidth,
    font=\footnotesize,
    axis y line*=left,
    ymin = 0, ymax = 200000,
    yticklabel style={anchor= near yticklabel},
    ylabel={Ridership \ref{pgfplots:plot1}},
    date coordinates in=x,
    xtick={ 
    2020-1-1,
    2020-2-1,
    2020-3-1,
    2020-4-1,
    2020-5-1,
    2020-6-1,
    2020-7-1},
    xticklabel style={rotate=45,anchor=near xticklabel},
    xticklabel={\pgfcalendarmonthshortname{\month} 1st},
    date ZERO=2009-08-18,
    legend style={at={(1,1)},anchor=north east,font=\scriptsize},
    enlarge x limits=true,
    ]
    \addplot+[red, mark=square] table [col sep=comma, x=date, y=board_count]
    {data/NashvilleWeeklyRidershipCovidcases.csv};
    \label{pgfplots:plot1}
\end{axis}
\begin{axis}[
    width=.8\linewidth,
    height=.3\linewidth,
    font=\footnotesize,
    axis y line*=right,
    axis x line=none,
    ymin = 0, ymax = 1600,
    ylabel={New Cases \ref{pgfplots:plot2}},
    xlabel=Date,
    date coordinates in=x,
    xtick={ 
    2020-1-1,
    2020-2-1,
    2020-3-1,
    2020-4-1,
    2020-5-1,
    2020-6-1,
    2020-7-1},
    xticklabel style={rotate=45,anchor=near xticklabel},
    xticklabel={\pgfcalendarmonthshortname{\month}},
    date ZERO=2009-08-18,
    enlarge x limits=true,
    ]
    \addplot+[blue, mark=o] table [col sep=comma, x=date, y=new_cases]
    {data/NashvilleWeeklyRidershipCovidcases.csv};
    \label{pgfplots:plot2},
    \addplot +[black, dashed, mark=none] coordinates {(2020-3-12, 0) (2020-3-12, 1560)};
    \addplot+[black, mark=x, mark options={scale=0}, ultra thick] coordinates {(2020-3-12, 100)} node[pin=170:{TN State of Emergency}]{};
    \addplot +[black, dashed, mark=none] coordinates {(2020-3-22, 0) (2020-3-22, 1560)};
    \addplot+[black, mark=x, mark options={scale=0}, ultra thick] coordinates {(2020-3-22, 400)} node[pin=170:{Nash. Stay at Home Order}]{};
    \addplot +[black, dashed, mark=none] coordinates {(2020-5-11, 0) (2020-5-11, 1600)};
    \addplot+[black, mark=x, mark options={scale=0}, ultra thick] coordinates {(2020-5-11, 800)} node[pin=170:{Phase 1}]{};
    \addplot +[black, dashed, mark=none] coordinates {(2020-5-25, 0) (2020-5-25, 1600)};
    \addplot+[black, mark=x, mark options={scale=0}, ultra thick] coordinates {(2020-5-25, 1000)} node[pin=170:{Phase 2}]{};
    \addplot +[black, dashed, mark=none] coordinates {(2020-6-29, 0) (2020-6-29, 1600)};
    \addplot+[black, mark=x, mark options={scale=0}, ultra thick] coordinates {(2020-6-29, 1000)} node[pin=95:{Phase 3}]{};
\end{axis}
\end{tikzpicture}
        \caption{Nashville}
        \label{subfig:Nashville_BoardingCovidcases}
    \end{subfigure}
    \begin{subfigure}{\linewidth}
        \centering
\pgfplotsset{scaled y ticks=false}
\begin{tikzpicture}
\begin{axis}[
    width=.8\linewidth,
    height=.3\linewidth,
    font=\footnotesize,
    axis y line*=left,
    ymin = 0, ymax = 30000,
    ylabel={Ridership \ref{pgfplots:plot11}},
    date coordinates in=x,
    xtick={ 
    2020-1-1,
    2020-2-1,
    2020-3-1,
    2020-4-1,
    2020-5-1,
    2020-6-1,
    2020-7-1},
    xticklabel style={rotate=45,anchor=near xticklabel},
    xticklabel={\pgfcalendarmonthshortname{\month} 1st},
    date ZERO=2009-08-18,
    enlarge x limits=true,
    ]
    \addplot+[red, mark=square] table [col sep=comma, x=date, y=board_count]
    {data/ChattanoogaWeeklyRidershipCovidcases.csv};
    \label{pgfplots:plot11},
\end{axis}
\begin{axis}[
    width=.8\linewidth,
    height=.3\linewidth,
    font=\footnotesize,
    axis y line*=right,
    axis x line=none,
    ymin = 0, ymax = 500,
    ylabel={New Cases \ref{pgfplots:plot22}},
    date coordinates in=x,
    xtick={ 
    2020-1-1,
    2020-2-1,
    2020-3-1,
    2020-4-1,
    2020-5-1,
    2020-6-1,
    2020-7-1},
    xticklabel style={rotate=45,anchor=near xticklabel},
    xticklabel={\pgfcalendarmonthshortname{\month}},
    date ZERO=2009-08-18,
    enlarge x limits=true,
    ]
    \addplot+[blue, mark=o] table [col sep=comma, x=date, y=new_cases]
    {data/ChattanoogaWeeklyRidershipCovidcases.csv};
    \label{pgfplots:plot22},
    \addplot +[black, dashed, mark=none] coordinates {(2020-3-12, 0) (2020-3-12, 1560)};
    \addplot+[black, mark=x, mark options={scale=0}, ultra thick] coordinates {(2020-3-12, 25)} node[pin=170:{TN State of Emergency}]{};
    \addplot +[black, dashed, mark=none] coordinates {(2020-3-22, 0) (2020-3-22, 1560)};
    \addplot+[black, mark=x, mark options={scale=0}, ultra thick] coordinates {(2020-3-22, 125)} node[pin=170:{TN Stay at Home Order}]{};
\end{axis}
\end{tikzpicture}
        \caption{Chattanooga}
        \label{subfig:Chattanooga_BoardingCovidcases}
    \end{subfigure}
    \begin{subfigure}{\linewidth}
        \centering
        \pgfplotstableread[col sep=comma]{data/NashvilleBaseline.csv}\NashvilleBaselineData

\pgfplotsset{scaled y ticks=false}
\begin{tikzpicture}
\begin{axis}[
    width=.8\linewidth,
    height=.3\linewidth,
    font=\footnotesize,
    ymin = -80, ymax = 40,
    ytick={-80, -60, -40, -20, 0, 20, 40},
    yticklabel style={anchor=near yticklabel},
    yticklabel={$\pgfmathprintnumber{\tick}\%$},
    ylabel={Change in Ridership},
    ylabel style = {align = center},
    ymajorgrids={true},
    date coordinates in=x,
    xtick={ 
    2020-1-1,
    2020-2-1,
    2020-3-1,
    2020-4-1,
    2020-5-1,
    2020-6-1,
     2020-7-1},
    xticklabel style={rotate=45,anchor=near xticklabel},
    xticklabel={\pgfcalendarmonthshortname{\month} 1st},
    date ZERO=2009-08-18,
	enlarge x limits=false,
	legend style={at={(1,1)},anchor=north east,font=\tiny}
    ]
    \addplot+[blue, mark=square] table [col sep=comma, x=date, y=change]
    {data/NashvilleBaseline.csv};
    \addlegendentry{Nashville},
    \addplot+[red, mark=o] table [col sep=comma, x=date, y=change]
    {data/ChattanoogaBaseline.csv};
    \addlegendentry{Chattanooga},
    \addplot+[black, mark=none] coordinates {(2020-1-1, 0) (2020-7-1, 0)};
\end{axis}
\end{tikzpicture}
        \caption{Change in ridership compared to 2019}
        \label{subfig:BoardingCovidcases2019}
    \end{subfigure}
    \caption{Weekly ridership compared to new COVID-19 cases per week for (a) Nashville and (b) Chattanooga. New COVID-19 cases are at the county level. Nashville is in Davidson County, Chattanooga is in Hamilton County. Phases 1, 2 and 3 in (a) are per Nashville's reopening plan provided by Nashville Metro. (c): Change in ridership compared to last year for Chattanooga and Nashville, TN from January through June 2020. Change in ridership was calculated by comparing weekly ridership to the baseline ridership from the same month in 2019.}
    \label{fig:BoardingCovidcases}
\end{figure*}
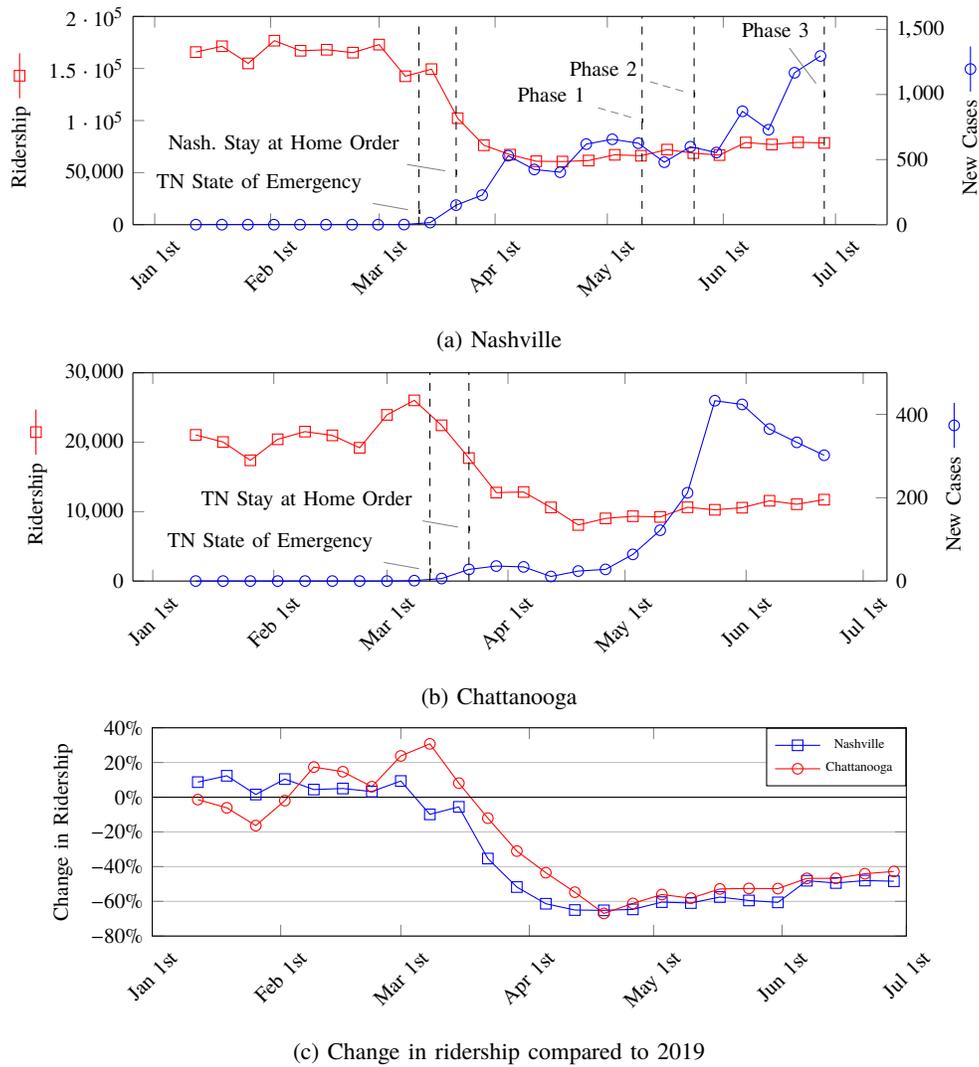

\else

\begin{figure*}[t!]
    \centering
    \begin{subfigure}{.8\linewidth}
        \centering
\pgfplotsset{scaled y ticks=false}
\begin{tikzpicture}
\begin{axis}[
    width=.8\linewidth,
    height=.3\linewidth,
    font=\footnotesize,
    axis y line*=left,
    ymin = 0, ymax = 200000,
    yticklabel style={anchor= near yticklabel},
    ylabel={Ridership \ref{pgfplots:plot1}},
    date coordinates in=x,
    xtick={ 
    2020-1-1,
    2020-2-1,
    2020-3-1,
    2020-4-1,
    2020-5-1,
    2020-6-1,
    2020-7-1},
    xticklabel style={rotate=45,anchor=near xticklabel},
    xticklabel={\pgfcalendarmonthshortname{\month} 1st},
    date ZERO=2009-08-18,
    legend style={at={(1,1)},anchor=north east,font=\scriptsize},
    enlarge x limits=true,
    ]
    \addplot+[red, mark=square] table [col sep=comma, x=date, y=board_count]
    {data/NashvilleWeeklyRidershipCovidcases.csv};
    \label{pgfplots:plot1}
\end{axis}
\begin{axis}[
    width=.8\linewidth,
    height=.3\linewidth,
    font=\footnotesize,
    axis y line*=right,
    axis x line=none,
    ymin = 0, ymax = 1600,
    ylabel={New Cases \ref{pgfplots:plot2}},
    xlabel=Date,
    date coordinates in=x,
    xtick={ 
    2020-1-1,
    2020-2-1,
    2020-3-1,
    2020-4-1,
    2020-5-1,
    2020-6-1,
    2020-7-1},
    xticklabel style={rotate=45,anchor=near xticklabel},
    xticklabel={\pgfcalendarmonthshortname{\month}},
    date ZERO=2009-08-18,
    enlarge x limits=true,
    ]
    \addplot+[blue, mark=o] table [col sep=comma, x=date, y=new_cases]
    {data/NashvilleWeeklyRidershipCovidcases.csv};
    \label{pgfplots:plot2},
    \addplot +[black, dashed, mark=none] coordinates {(2020-3-12, 0) (2020-3-12, 1560)};
    \addplot+[black, mark=x, mark options={scale=0}, ultra thick] coordinates {(2020-3-12, 100)} node[pin=170:{TN State of Emergency}]{};
    \addplot +[black, dashed, mark=none] coordinates {(2020-3-22, 0) (2020-3-22, 1560)};
    \addplot+[black, mark=x, mark options={scale=0}, ultra thick] coordinates {(2020-3-22, 400)} node[pin=170:{Nash. Stay at Home Order}]{};
    \addplot +[black, dashed, mark=none] coordinates {(2020-5-11, 0) (2020-5-11, 1600)};
    \addplot+[black, mark=x, mark options={scale=0}, ultra thick] coordinates {(2020-5-11, 800)} node[pin=170:{Phase 1}]{};
    \addplot +[black, dashed, mark=none] coordinates {(2020-5-25, 0) (2020-5-25, 1600)};
    \addplot+[black, mark=x, mark options={scale=0}, ultra thick] coordinates {(2020-5-25, 1000)} node[pin=170:{Phase 2}]{};
    \addplot +[black, dashed, mark=none] coordinates {(2020-6-29, 0) (2020-6-29, 1600)};
    \addplot+[black, mark=x, mark options={scale=0}, ultra thick] coordinates {(2020-6-29, 1000)} node[pin=95:{Phase 3}]{};
\end{axis}
\end{tikzpicture}
        \caption{Nashville}
        \label{subfig:Nashville_BoardingCovidcases}
    \end{subfigure}
    \begin{subfigure}{.8\linewidth}
        \centering
\pgfplotsset{scaled y ticks=false}
\begin{tikzpicture}
\begin{axis}[
    width=.8\linewidth,
    height=.3\linewidth,
    font=\footnotesize,
    axis y line*=left,
    ymin = 0, ymax = 30000,
    ylabel={Ridership \ref{pgfplots:plot11}},
    date coordinates in=x,
    xtick={ 
    2020-1-1,
    2020-2-1,
    2020-3-1,
    2020-4-1,
    2020-5-1,
    2020-6-1,
    2020-7-1},
    xticklabel style={rotate=45,anchor=near xticklabel},
    xticklabel={\pgfcalendarmonthshortname{\month} 1st},
    date ZERO=2009-08-18,
    enlarge x limits=true,
    ]
    \addplot+[red, mark=square] table [col sep=comma, x=date, y=board_count]
    {data/ChattanoogaWeeklyRidershipCovidcases.csv};
    \label{pgfplots:plot11},
\end{axis}
\begin{axis}[
    width=.8\linewidth,
    height=.3\linewidth,
    font=\footnotesize,
    axis y line*=right,
    axis x line=none,
    ymin = 0, ymax = 500,
    ylabel={New Cases \ref{pgfplots:plot22}},
    date coordinates in=x,
    xtick={ 
    2020-1-1,
    2020-2-1,
    2020-3-1,
    2020-4-1,
    2020-5-1,
    2020-6-1,
    2020-7-1},
    xticklabel style={rotate=45,anchor=near xticklabel},
    xticklabel={\pgfcalendarmonthshortname{\month}},
    date ZERO=2009-08-18,
    enlarge x limits=true,
    ]
    \addplot+[blue, mark=o] table [col sep=comma, x=date, y=new_cases]
    {data/ChattanoogaWeeklyRidershipCovidcases.csv};
    \label{pgfplots:plot22},
    \addplot +[black, dashed, mark=none] coordinates {(2020-3-12, 0) (2020-3-12, 1560)};
    \addplot+[black, mark=x, mark options={scale=0}, ultra thick] coordinates {(2020-3-12, 25)} node[pin=170:{TN State of Emergency}]{};
    \addplot +[black, dashed, mark=none] coordinates {(2020-3-22, 0) (2020-3-22, 1560)};
    \addplot+[black, mark=x, mark options={scale=0}, ultra thick] coordinates {(2020-3-22, 125)} node[pin=170:{TN Stay at Home Order}]{};
\end{axis}
\end{tikzpicture}
        \caption{Chattanooga}
        \label{subfig:Chattanooga_BoardingCovidcases}
    \end{subfigure}
    \begin{subfigure}{.8\linewidth}
        \centering
        \pgfplotstableread[col sep=comma]{data/NashvilleBaseline.csv}\NashvilleBaselineData

\pgfplotsset{scaled y ticks=false}
\begin{tikzpicture}
\begin{axis}[
    width=.8\linewidth,
    height=.3\linewidth,
    font=\footnotesize,
    ymin = -80, ymax = 40,
    ytick={-80, -60, -40, -20, 0, 20, 40},
    yticklabel style={anchor=near yticklabel},
    yticklabel={$\pgfmathprintnumber{\tick}\%$},
    ylabel={Change in Ridership},
    ylabel style = {align = center},
    ymajorgrids={true},
    date coordinates in=x,
    xtick={ 
    2020-1-1,
    2020-2-1,
    2020-3-1,
    2020-4-1,
    2020-5-1,
    2020-6-1,
     2020-7-1},
    xticklabel style={rotate=45,anchor=near xticklabel},
    xticklabel={\pgfcalendarmonthshortname{\month} 1st},
    date ZERO=2009-08-18,
	enlarge x limits=false,
	legend style={at={(1,1)},anchor=north east,font=\tiny}
    ]
    \addplot+[blue, mark=square] table [col sep=comma, x=date, y=change]
    {data/NashvilleBaseline.csv};
    \addlegendentry{Nashville},
    \addplot+[red, mark=o] table [col sep=comma, x=date, y=change]
    {data/ChattanoogaBaseline.csv};
    \addlegendentry{Chattanooga},
    \addplot+[black, mark=none] coordinates {(2020-1-1, 0) (2020-7-1, 0)};
\end{axis}
\end{tikzpicture}
        \caption{Change in ridership compared to 2019}
        \label{subfig:BoardingCovidcases2019}
    \end{subfigure}
    \caption{Weekly ridership compared to new COVID-19 cases per week for (a) Nashville and (b) Chattanooga. New COVID-19 cases are at the county level. Nashville is in Davidson County, Chattanooga is in Hamilton County. Phases 1, 2 and 3 in (a) are per Nashville's reopening plan provided by Nashville Metro. (c): Change in ridership compared to last year for Chattanooga and Nashville, TN from January through June 2020. Change in ridership was calculated by comparing weekly ridership to the baseline ridership from the same month in 2019.}
    \label{fig:BoardingCovidcases}
\end{figure*}
\fi
\else
  
  \subsection{Impact of COVID-19 on city-wide ridership}\label{subsec:impacts of covid}
\fi

The fundamental question in this section is to what degree has COVID-19 changed ridership and what effects do these changes have on transit operations. \cref{subfig:Nashville_BoardingCovidcases} and \cref{subfig:Chattanooga_BoardingCovidcases} show weekly total ridership and weekly new COVID-19 cases in Nashville and Chattanooga respectively. \cref{subfig:BoardingCovidcases2019} shows drop in ridership for Nashville and Chattanooga compared to a baseline. The baseline was calculated by taking the average weekly ridership for the corresponding month in 2019 for both cities.

As shown in \cref{subfig:Nashville_BoardingCovidcases}, Nashville public transit ridership started to decline on the week of March 1st which corresponds with the first known COVID-19 case in Tennessee on March 5th and the Tennessee State of Emergency Order on March 12. Perhaps more importantly there was a major tornado in Nashville on March 3rd \cite{tornado} which helps explain the initial decline in ridership at this time. Ridership remained constant for a week before a significant decline started during the week of March 22nd when the Nashville Safer at Home Order was put into effect on March 22, 2020. Nashville ultimately reached a low of 60,620 riders on the week of April 19, a 66.9\% reduction from the average ridership in April of 2019 as shown in
\cref{subfig:BoardingCovidcases2019}.
From late April to July 1st ridership stabilized. By the week of June 28th ridership in Nashville has recovered 22.7\% from the low in April. Chattanooga's steep decline started the week of March 5th before hitting a low also on the week of April 19 with a low of 8,077 weekly riders as shown in \cref{subfig:Chattanooga_BoardingCovidcases}. Compared to the 2019 baseline, Chattanooga had a 65.1\% in ridership on April 19 \cref{subfig:BoardingCovidcases2019}. Ultimately Chattanooga ridership recovered to 11,725 riders the week of Jun 28th, an increase of 45.2\% from the low of April 19-25. 

Ultimately, both cities saw a rapid decline in ridership from early March to late April before ridership stabilized through the end of June. To characterize these findings, we refer to January through February as pre-COVID operations and starting in late April a new normal post-COVID operations stabilized at approximately 60\% reduction in ridership compared to the previous year for both cities. 


\subsection{Route level investigation}

\if 1\mode
\begin{figure}[t]
    \centering
    \begin{subfigure}{\linewidth}
        \centering
        \pgfplotstableread[col sep=comma]{data/NashvilleMonthlyAverageRidershipbyTopFiveRoute.csv}\RouteWeekly


\begin{tikzpicture}[font=\scriptsize]
    \begin{axis}
        [
	       ylabel=Ridership,
            ylabel style = {align = center},
            	grid=major,
            	SmallBarPlot,
            xtick = data,
            xticklabels={January,February,March,April,May,June,July},
            tick label style={font=\scriptsize}, 
            legend pos=north east,
            width=\linewidth,
            height=4.5cm,
            xmin = -0.5, xmax = 6,
            legend columns=3, 
            legend style={at={(1,1)},anchor=north east,font=\tiny}
        ]
            \addplot [red!60, smooth, mark=*] table [x expr=\coordindex, y=Route_22] {\RouteWeekly};
            \addlegendentry{Route 22},
            
            \addplot [blue!50, smooth, mark=diamond*] table [x expr=\coordindex, y=Route_50] {\RouteWeekly};
            \addlegendentry{Route 50},
            
            \addplot [orange!90, smooth, mark=triangle*] table [x expr=\coordindex, y=Route_52] {\RouteWeekly};
            \addlegendentry{Route 52},
            
            \addplot [green!60, smooth, mark=square*] table [x expr=\coordindex, y=Route_55] {\RouteWeekly};
            \addlegendentry{Route 55},
            \addplot [black, smooth, mark=pentagon*] table [x expr=\coordindex, y=Route_56] {\RouteWeekly};
            \addlegendentry{Route 56}
    \end{axis}
\end{tikzpicture}

        \caption{}
        \label{subfig:Nashville_TopFiveRouteWeekly}
    \end{subfigure}
    \begin{subfigure}{\linewidth}
        \centering
        \pgfplotstableread[col sep=comma]{data/ChattanoogaMonthlyAverageRidershipbyTopFiveRoute.csv}\RouteWeekly


\begin{tikzpicture}[font=\scriptsize]
    \begin{axis}
        [
	       ylabel=Ridership,
            ylabel style = {align = center},
            	grid=major,
            	SmallBarPlot,
            xtick = data,
            xticklabels={January,February,March,April,May,June,July},
            tick label style={font=\scriptsize}, 
            legend pos=north east,
            width=\linewidth,
            height=4.5cm,
            xmin = -0.5, xmax =6,
            legend columns=3, 
            legend style={at={(1,1)},anchor=north east,font=\tiny}
        ]
            
            \addplot [red!60, smooth, mark=*] table [x expr=\coordindex, y=Route_1] {\RouteWeekly};
            \addlegendentry{Route 1},
            \addplot [blue!50, smooth, mark=diamond*] table [x expr=\coordindex, y=Route_4] {\RouteWeekly};
            \addlegendentry{Route 4},
            \addplot [orange!90, smooth, mark=triangle*] table [x expr=\coordindex, y=Route_9] {\RouteWeekly};
            \addlegendentry{Route 9},
            \addplot [green!40, smooth, mark=square*] table [x expr=\coordindex, y=Route_10] {\RouteWeekly};
            \addlegendentry{Route 10},
            \addplot [black, smooth, mark=pentagon*] table [x expr=\coordindex, y=Route_14] {\RouteWeekly};
            \addlegendentry{Route 14}
    \end{axis}
\end{tikzpicture}

        \caption{}
        \label{subfig:Chattanooga_TopFiveRouteWeekly}
    \end{subfigure}
    \caption{Average weekly ridership per month for the 5 most popular routes in (a) Nashville and (b) Chattanooga in 2020. Ridership trends follow similar patterns to the demand patterns across all routes in \cref{fig:BoardingCovidcases}.}
    \label{fig:TopFiveRouteWeekly}
\end{figure}

\else
\begin{figure}[h!]
    \centering
    \begin{subfigure}{.85\textwidth}
        \centering
        \pgfplotstableread[col sep=comma]{data/NashvilleMonthlyAverageRidershipbyTopFiveRoute.csv}\RouteWeekly


\begin{tikzpicture}[font=\scriptsize]
    \begin{axis}
        [
	       ylabel=Ridership,
            ylabel style = {align = center},
            	grid=major,
            	SmallBarPlot,
            xtick = data,
            xticklabels={January,February,March,April,May,June,July},
            tick label style={font=\scriptsize}, 
            legend pos=north east,
            width=\linewidth,
            height=4.5cm,
            xmin = -0.5, xmax = 6,
            legend columns=3, 
            legend style={at={(1,1)},anchor=north east,font=\tiny}
        ]
            \addplot [red!60, smooth, mark=*] table [x expr=\coordindex, y=Route_22] {\RouteWeekly};
            \addlegendentry{Route 22},
            
            \addplot [blue!50, smooth, mark=diamond*] table [x expr=\coordindex, y=Route_50] {\RouteWeekly};
            \addlegendentry{Route 50},
            
            \addplot [orange!90, smooth, mark=triangle*] table [x expr=\coordindex, y=Route_52] {\RouteWeekly};
            \addlegendentry{Route 52},
            
            \addplot [green!60, smooth, mark=square*] table [x expr=\coordindex, y=Route_55] {\RouteWeekly};
            \addlegendentry{Route 55},
            \addplot [black, smooth, mark=pentagon*] table [x expr=\coordindex, y=Route_56] {\RouteWeekly};
            \addlegendentry{Route 56}
    \end{axis}
\end{tikzpicture}

        \caption{}
        \label{subfig:Nashville_TopFiveRouteWeekly}
    \end{subfigure}
    \begin{subfigure}{.85\textwidth}
        \centering
        \pgfplotstableread[col sep=comma]{data/ChattanoogaMonthlyAverageRidershipbyTopFiveRoute.csv}\RouteWeekly


\begin{tikzpicture}[font=\scriptsize]
    \begin{axis}
        [
	       ylabel=Ridership,
            ylabel style = {align = center},
            	grid=major,
            	SmallBarPlot,
            xtick = data,
            xticklabels={January,February,March,April,May,June,July},
            tick label style={font=\scriptsize}, 
            legend pos=north east,
            width=\linewidth,
            height=4.5cm,
            xmin = -0.5, xmax =6,
            legend columns=3, 
            legend style={at={(1,1)},anchor=north east,font=\tiny}
        ]
            
            \addplot [red!60, smooth, mark=*] table [x expr=\coordindex, y=Route_1] {\RouteWeekly};
            \addlegendentry{Route 1},
            \addplot [blue!50, smooth, mark=diamond*] table [x expr=\coordindex, y=Route_4] {\RouteWeekly};
            \addlegendentry{Route 4},
            \addplot [orange!90, smooth, mark=triangle*] table [x expr=\coordindex, y=Route_9] {\RouteWeekly};
            \addlegendentry{Route 9},
            \addplot [green!40, smooth, mark=square*] table [x expr=\coordindex, y=Route_10] {\RouteWeekly};
            \addlegendentry{Route 10},
            \addplot [black, smooth, mark=pentagon*] table [x expr=\coordindex, y=Route_14] {\RouteWeekly};
            \addlegendentry{Route 14}
    \end{axis}
\end{tikzpicture}

        \caption{}
        \label{subfig:Chattanooga_TopFiveRouteWeekly}
    \end{subfigure}
    \caption{Average weekly ridership per month for the 5 most popular routes in (a) Nashville and (b) Chattanooga in 2020. Ridership trends follow similar patterns to the demand patterns across all routes in \cref{fig:BoardingCovidcases}.}
    \label{fig:TopFiveRouteWeekly}
\end{figure}
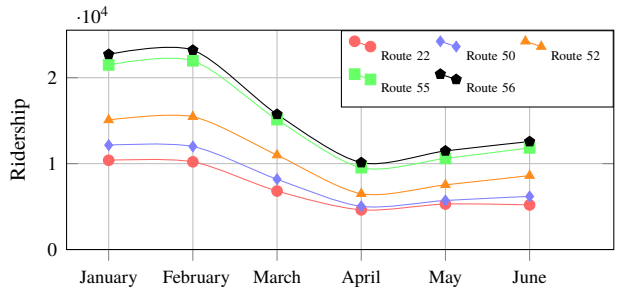
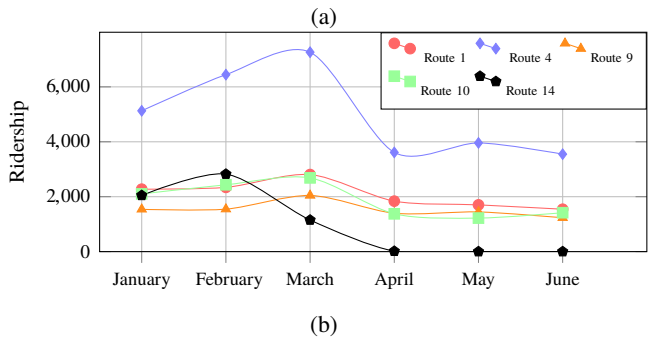
\fi

\cref{subfig:Nashville_TopFiveRouteWeekly} and \cref{subfig:Chattanooga_TopFiveRouteWeekly} show the monthly ridership distribution on the top 5 routes for the city of Nashville and Chattanooga respectively. We see similar trends to the aggregated ridership analysis in the previous section. For both cities ridership decreased rapidly before stabilizing in April. In Nashville however we see a greater rebound between April to June than in Chattanooga. The rebound in Nashville corresponds loosely with Phase 2 reopening.

An important note is that for Chattanooga route 14 is one of the most used routes however it is unique in that it is a free shuttle service to the University of Tennessee, Chattanooga. When Universities went online in March route 14 initially continued operating on its regular Saturday schedule. Due to the drastic demand reduction during this time Chattanooga ultimately stopped the service entirely in April. 

From this section we see that the most populated routes follow a similar trajectory and magnitude of ridership drop as the fixed-line transit system overall. Therefore a more detailed temporal and spatial analysis is outlined in the following sections of this paper.

\subsection{Temporal analysis of transit usage and rider behavior}

\if 1\mode

\begin{figure}[t]
    \centering
    \begin{subfigure}{\linewidth}
        \centering
        \pgfplotstableread[col sep=comma]{data/Nashville2020PrePostCovidWeekOfDayAverageRiderShip.csv}\PrePostCOVIDWeeklyRidership


\begin{tikzpicture}[font=\scriptsize]
    \begin{axis}
        [
	       ylabel=Ridership,
            ylabel style = {align = center},
            	grid=major,
            	ybar,
            xtick = data,
            xticklabels=
            {Monday,Tuesday, Wednesday, Thursday, Friday, Saturday, Sunday},
            xticklabel style={rotate=45,anchor=near xticklabel},
            tick label style={font=\scriptsize}, 
            legend pos=north east,
            width=\linewidth,
            height=3.5cm,
            xmin = -1, xmax = 7,
            legend style={at={(1,1)},anchor=north east,font=\scriptsize}
        ]
            \addplot [RedBigBars] table [x expr=\coordindex, y=JanFebRIDERSHIP] {\PrePostCOVIDWeeklyRidership};
            \addlegendentry{Jan-Feb},
            \addplot [BlueBigBars] table [x expr=\coordindex, y=MayJunRIDERSHIP] {\PrePostCOVIDWeeklyRidership};
            \addlegendentry{May-June}
    \end{axis}
\end{tikzpicture}

        \caption{}
        \label{subfig:Nashville_PrePostCovidWeeklyAverage}
    \end{subfigure}
    \begin{subfigure}{\linewidth}
        \centering
        \pgfplotstableread[col sep=comma]{data/Chattanooga2020PrePostCovidDayOfWeekAverageRiderShip.csv}\PrePostCOVIDWeeklyRidership


\begin{tikzpicture}[font=\scriptsize]
    \begin{axis}
        [
	       ylabel=Ridership,
            ylabel style = {align = center},
            	grid=major,
            	ybar,
            xtick = data,
            xticklabels=
            {Monday,Tuesday, Wednesday, Thursday, Friday, Saturday, Sunday},
            xticklabel style={rotate=45,anchor=near xticklabel},
            tick label style={font=\scriptsize}, 
            legend pos=north east,
            width=\linewidth,
            height=3.5cm,
            xmin = -1, xmax = 7,
            legend style={at={(1,1)},anchor=north east,font=\scriptsize}
        ]
            \addplot [RedBigBars] table [x expr=\coordindex, y=JanFebRIDERSHIP] {\PrePostCOVIDWeeklyRidership};
            \addlegendentry{Jan-Feb},
            \addplot [BlueBigBars] table [x expr=\coordindex, y=MayJunRIDERSHIP] {\PrePostCOVIDWeeklyRidership};
            \addlegendentry{May-June}
    \end{axis}
\end{tikzpicture}

        \caption{}
        \label{subfig:Chattanooga_PrePostCovidWeeklyAverage}
    \end{subfigure}
    \caption{Daily average ridership for January--February and May--June 2020 for (a) Nashville and (b) Chattanooga. January--February represents baseline pre-COVID ridership levels in 2020 while May--June represents ridership after it stabilized post-COVID.}
    \vspace{-1em}
\end{figure}
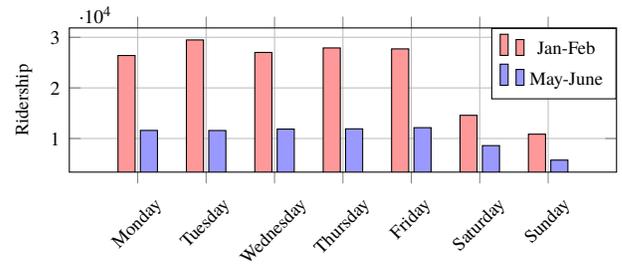
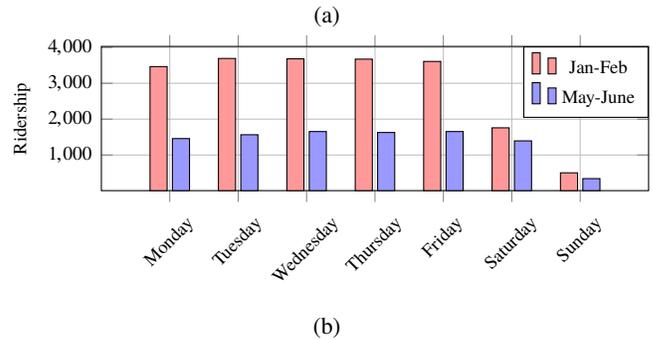

\else
\begin{figure}[]
    \centering
    \begin{subfigure}{.85\textwidth}
        \centering
        \pgfplotstableread[col sep=comma]{data/Nashville2020PrePostCovidWeekOfDayAverageRiderShip.csv}\PrePostCOVIDWeeklyRidership


\begin{tikzpicture}[font=\scriptsize]
    \begin{axis}
        [
	       ylabel=Ridership,
            ylabel style = {align = center},
            	grid=major,
            	ybar,
            xtick = data,
            xticklabels=
            {Monday,Tuesday, Wednesday, Thursday, Friday, Saturday, Sunday},
            xticklabel style={rotate=45,anchor=near xticklabel},
            tick label style={font=\scriptsize}, 
            legend pos=north east,
            width=\linewidth,
            height=3.5cm,
            xmin = -1, xmax = 7,
            legend style={at={(1,1)},anchor=north east,font=\scriptsize}
        ]
            \addplot [RedBigBars] table [x expr=\coordindex, y=JanFebRIDERSHIP] {\PrePostCOVIDWeeklyRidership};
            \addlegendentry{Jan-Feb},
            \addplot [BlueBigBars] table [x expr=\coordindex, y=MayJunRIDERSHIP] {\PrePostCOVIDWeeklyRidership};
            \addlegendentry{May-June}
    \end{axis}
\end{tikzpicture}

        \caption{}
        \label{subfig:Nashville_PrePostCovidWeeklyAverage}
    \end{subfigure}
    \begin{subfigure}{.85\textwidth}
        \centering
        \pgfplotstableread[col sep=comma]{data/Chattanooga2020PrePostCovidDayOfWeekAverageRiderShip.csv}\PrePostCOVIDWeeklyRidership


\begin{tikzpicture}[font=\scriptsize]
    \begin{axis}
        [
	       ylabel=Ridership,
            ylabel style = {align = center},
            	grid=major,
            	ybar,
            xtick = data,
            xticklabels=
            {Monday,Tuesday, Wednesday, Thursday, Friday, Saturday, Sunday},
            xticklabel style={rotate=45,anchor=near xticklabel},
            tick label style={font=\scriptsize}, 
            legend pos=north east,
            width=\linewidth,
            height=3.5cm,
            xmin = -1, xmax = 7,
            legend style={at={(1,1)},anchor=north east,font=\scriptsize}
        ]
            \addplot [RedBigBars] table [x expr=\coordindex, y=JanFebRIDERSHIP] {\PrePostCOVIDWeeklyRidership};
            \addlegendentry{Jan-Feb},
            \addplot [BlueBigBars] table [x expr=\coordindex, y=MayJunRIDERSHIP] {\PrePostCOVIDWeeklyRidership};
            \addlegendentry{May-June}
    \end{axis}
\end{tikzpicture}

        \caption{}
        \label{subfig:Chattanooga_PrePostCovidWeeklyAverage}
    \end{subfigure}
    \caption{Daily average ridership for January--February and May--June 2020 for (a) Nashville and (b) Chattanooga. January--February represents baseline pre-COVID ridership levels in 2020 while May--June represents ridership after it stabilized post-COVID.}
    \vspace{-1em}
\end{figure}
\fi
\if 1\mode

\begin{figure}[t]
    \centering
    \begin{subfigure}{\linewidth}
        \centering
        \pgfplotstableread[col sep=comma]{data/Nashville2020PrePostCovidHourlyAverageRiderShip.csv}\PrePostCOVIDhourlyRidership


\centering
\begin{tikzpicture}[font=\scriptsize]
    \begin{axis}
        [
	       ylabel=Ridership,
            ylabel style = {align = center},
            	grid=major,
            	ybar,
            xtick = data,
            xticklabels from table={\PrePostCOVIDhourlyRidership}{Hour},
            tick label style={font=\scriptsize}, 
            legend pos=north east,
            width=\linewidth,
            height=3.5cm,
            xmin = -1, xmax = 21,
            legend style={at={(1,1)},anchor=north east,font=\scriptsize}
        ]
            \addplot [RedBigBars] table [x expr=\coordindex, y=JanFebRIDERSHIP] {\PrePostCOVIDhourlyRidership};
            \addlegendentry{Jan-Feb},
            \addplot [BlueBigBars] table [x expr=\coordindex, y=MayJunRIDERSHIP] {\PrePostCOVIDhourlyRidership};
            \addlegendentry{May-Jun}
    \end{axis}
\end{tikzpicture}

        \caption{}
        \label{subfig:Nashville_PrePostCovidHourlyAverage}
    \end{subfigure}
    \begin{subfigure}{\linewidth}
        \centering
        \pgfplotstableread[col sep=comma]{data/Chattanooga2020PrePostCovidHourlyAverageRiderShip.csv}\PrePostCOVIDhourlyRidership


\centering
\begin{tikzpicture}[font=\scriptsize]
    \begin{axis}
        [
            xlabel=Time of Day (Hour),
	       ylabel=Ridership,
            ylabel style = {align = center},
            	grid=major,
            	ybar,
            xtick = data,
            xticklabels from table={\PrePostCOVIDhourlyRidership}{Hour},
            tick label style={font=\scriptsize}, 
            legend pos=north east,
            width=\linewidth,
            height=3.5cm,
            xmin = -1, xmax = 21,
            legend style={at={(1,1)},anchor=north east,font=\scriptsize}
        ]
            \addplot [RedBigBars] table [x expr=\coordindex, y=JanFebRIDERSHIP] {\PrePostCOVIDhourlyRidership};
            \addlegendentry{Jan-Feb},
            \addplot [BlueBigBars] table [x expr=\coordindex, y=MayJunRIDERSHIP] {\PrePostCOVIDhourlyRidership};
            \addlegendentry{May-June}
    \end{axis}
\end{tikzpicture}

        \caption{}
        \label{subfig:Chattanooga_PrePostCovidHourlyAverage}
    \end{subfigure}
    \caption{Average ridership per hour of day for January--February and May--June 2020 for (a) Nashville and (b) Chattanooga. January--February represents baseline pre-COVID ridership levels in 2020 while May--June represents ridership after it stabilized post-COVID.}
   
\end{figure}

\else
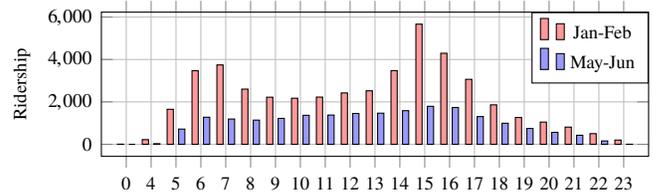
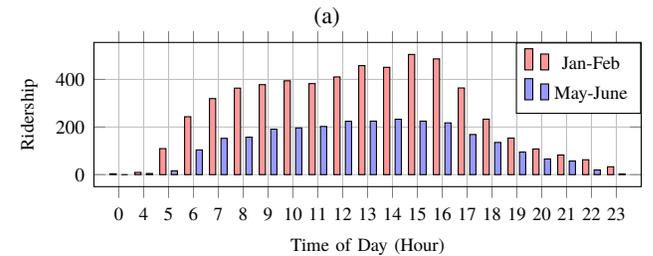
\begin{figure}[]
    \centering
    \begin{subfigure}{.9\textwidth}
        \centering
        \pgfplotstableread[col sep=comma]{data/Nashville2020PrePostCovidHourlyAverageRiderShip.csv}\PrePostCOVIDhourlyRidership


\centering
\begin{tikzpicture}[font=\scriptsize]
    \begin{axis}
        [
	       ylabel=Ridership,
            ylabel style = {align = center},
            	grid=major,
            	ybar,
            xtick = data,
            xticklabels from table={\PrePostCOVIDhourlyRidership}{Hour},
            tick label style={font=\scriptsize}, 
            legend pos=north east,
            width=\linewidth,
            height=3.5cm,
            xmin = -1, xmax = 21,
            legend style={at={(1,1)},anchor=north east,font=\scriptsize}
        ]
            \addplot [RedBigBars] table [x expr=\coordindex, y=JanFebRIDERSHIP] {\PrePostCOVIDhourlyRidership};
            \addlegendentry{Jan-Feb},
            \addplot [BlueBigBars] table [x expr=\coordindex, y=MayJunRIDERSHIP] {\PrePostCOVIDhourlyRidership};
            \addlegendentry{May-Jun}
    \end{axis}
\end{tikzpicture}

        \caption{}
        \label{subfig:Nashville_PrePostCovidHourlyAverage}
    \end{subfigure}
    \begin{subfigure}{.9\textwidth}
        \centering
        \pgfplotstableread[col sep=comma]{data/Chattanooga2020PrePostCovidHourlyAverageRiderShip.csv}\PrePostCOVIDhourlyRidership


\centering
\begin{tikzpicture}[font=\scriptsize]
    \begin{axis}
        [
            xlabel=Time of Day (Hour),
	       ylabel=Ridership,
            ylabel style = {align = center},
            	grid=major,
            	ybar,
            xtick = data,
            xticklabels from table={\PrePostCOVIDhourlyRidership}{Hour},
            tick label style={font=\scriptsize}, 
            legend pos=north east,
            width=\linewidth,
            height=3.5cm,
            xmin = -1, xmax = 21,
            legend style={at={(1,1)},anchor=north east,font=\scriptsize}
        ]
            \addplot [RedBigBars] table [x expr=\coordindex, y=JanFebRIDERSHIP] {\PrePostCOVIDhourlyRidership};
            \addlegendentry{Jan-Feb},
            \addplot [BlueBigBars] table [x expr=\coordindex, y=MayJunRIDERSHIP] {\PrePostCOVIDhourlyRidership};
            \addlegendentry{May-June}
    \end{axis}
\end{tikzpicture}

        \caption{}
        \label{subfig:Chattanooga_PrePostCovidHourlyAverage}
    \end{subfigure}
    \caption{Average ridership per hour of day for January--February and May--June 2020 for (a) Nashville and (b) Chattanooga. January--February represents baseline pre-COVID ridership levels in 2020 while May--June represents ridership after it stabilized post-COVID.}
   
\end{figure}
\fi

Here we investigate temporal changes in ridership between pre-COVID and post-COVID operations. As discussed in Section \ref{subsec:impacts of covid}, for both cities normal operations spanned from January 1st to the end of February and after a rapid drop ridership stabilized in mid-to-late April. Therefore in this Section we use January-February to represent pre-COVID operations and May-June to represent post-COVID operations. In \cref{subfig:Nashville_PrePostCovidWeeklyAverage} and \cref{subfig:Chattanooga_PrePostCovidWeeklyAverage}, we see the ridership distribution of Nashville and Chattanooga for each day of the week before COVID-19 and after COVID-19. In both cities the drop in ridership on the weekends is less than weekdays with Chattanooga only seeing a 20\% decrease in ridership on Saturdays and a 32\% decrease on Sundays compared to an average of 56\% on weekdays. Nashville saw a 41\% decrease in ridership on Saturdays and a 47\% decrease on Sundays compared to an average of 57\% decrease for weekdays. 

\cref{subfig:Nashville_PrePostCovidHourlyAverage} and \cref{subfig:Chattanooga_PrePostCovidHourlyAverage} show ridership in January-February compared to May-June per hour of the day. We can see that the biggest drops in ridership occur during morning rush and evening rush. This is highlighted in Nashville where morning rush (5:00AM-9:00AM) saw a 64\% change in ridership and evening rush (3:00PM-6:00PM) saw a 62\% decrease compared to 42\% change between 9:00AM and 3:00PM. This discrepancy was not as pronounced with Chattanooga where there was a 62\% and 56\% decrease in ridership for morning and evening rush respectively compared to a 53\% between 9:00AM and 3:00PM.

As we can see in this section, the biggest declines in ridership were on weekdays during morning and evening commuting times. The declines continued after the Nashville and Tennessee Stay at Home orders expired showing a persistent shift towards alternative work options throughout the COVID-19 pandemic. This phenomena was however more apparent in Nashville than Chattanooga.

\subsection{Spatial analysis of transit usage and rider behavior}\label{subsec:spatial}
\if 0\mode
\begin{figure*}

\centering
\begin{subfigure}{.4\textwidth}
  \centering
  \includegraphics[width=\linewidth]{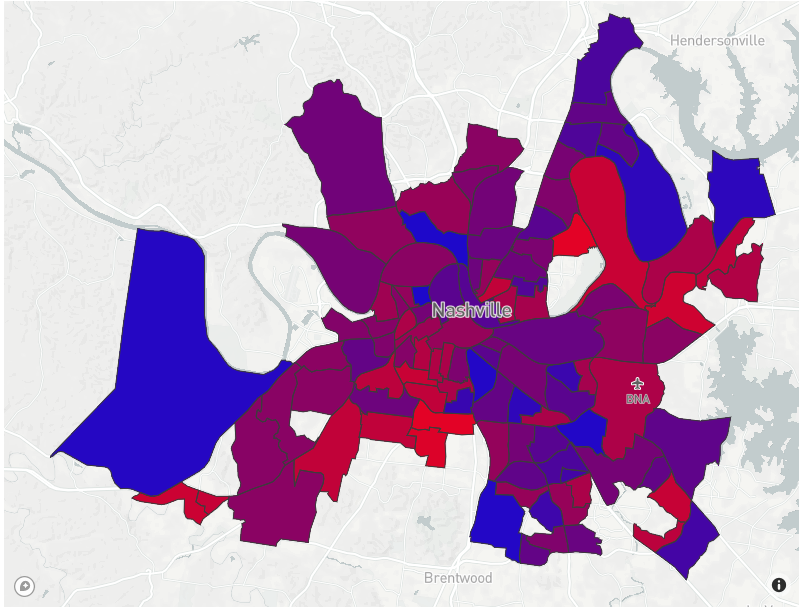}
  \label{fig:sub1}
\end{subfigure}%
\begin{subfigure}{.4\textwidth}
  \centering
  \includegraphics[width=\linewidth]{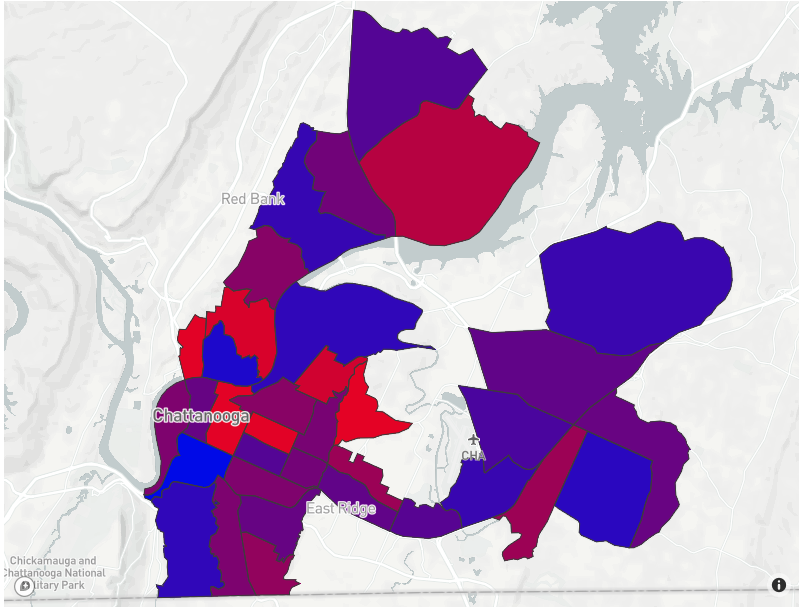}
  \label{fig:sub2}
\end{subfigure}
\begin{subfigure}{.1\textwidth}
  \centering
  \includegraphics[width=.8\linewidth]{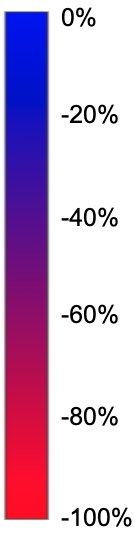}
  {\caption*{.}}
\end{subfigure}
\caption{Change in ridership between pre-COVID (January--February) and post-COVID (May--June) 2020 per census tract for (left) Nashville and (right) Chattanooga. }
\label{fig:ChoroFig}
\end{figure*}

\else

\begin{figure*}[]
\centering
\begin{subfigure}{.4\textwidth}
  \centering
  \includegraphics[width=\linewidth]{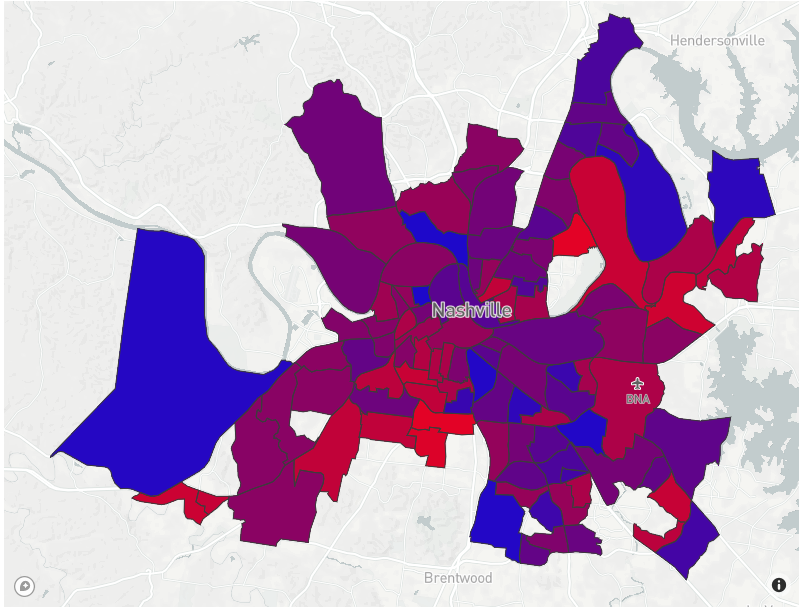}
  \label{fig:sub1}
\end{subfigure}%
\begin{subfigure}{.4\textwidth}
  \centering
  \includegraphics[width=\linewidth]{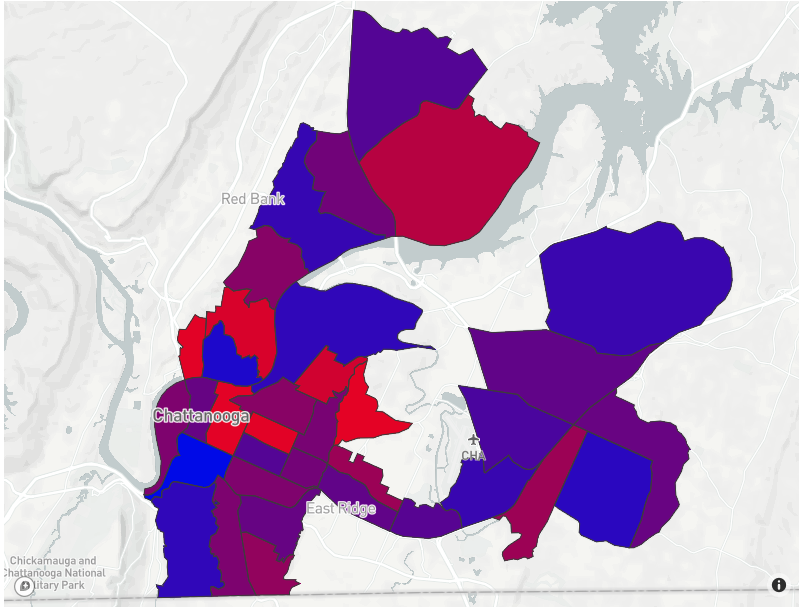}
  \label{fig:sub2}
\end{subfigure}
\begin{subfigure}{.1\textwidth}
  \centering
  \includegraphics[width=.8\linewidth]{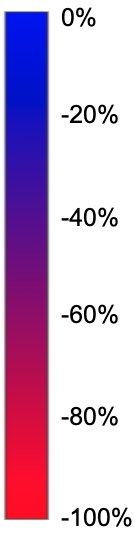}
  {\caption*{.}}
\end{subfigure}
\caption{Change in ridership between pre-COVID (January--February) and post-COVID (May--June) 2020 per census tract for (left) Nashville and (right) Chattanooga. }
\label{fig:ChoroFig}
\end{figure*}
\fi

In this section we look at spatial variation in ridership. Each boarding was mapped to a corresponding 2010 census tract in which that boarding occurred. \cref{fig:ChoroFig} shows the percent decrease in ridership between pre-COVID (January-February) and post-COVID (May-June) operations per census tract. As shown, change in ridership was not uniformly spaced throughout either city. Both cities see significant decreases downtown, most likely due to workers working remotely. This was most visible in Chattanooga where ridership decreased by up to 81\%. Chattanooga also saw a significant decrease in ridership in the census tract that contains the University of Tennessee, Chattanooga reflecting the University's decision to suspend in-person operations and CARTA's subsequent cancellation of the free shuttle servicing this region. While the same patterns are present in Nashville, change in ridership was more uniform likely due to the density of Nashville's downtown region. Nashville saw significant decrease in ridership to areas heavily dependent on retail and shopping which includes a 87\% drop to Opry Mills and a 86\% drop to Green Hills.

\subsection{Socio-economic analysis}\label{subsec:socio}

\if 1\mode
\begin{figure}[]
    \centering
\pgfplotsset{scaled y ticks=false}
\begin{tikzpicture}

\begin{axis}[
    width=\columnwidth,
    height=5cm,
    font=\footnotesize,
    ymin = -100, ymax = 40,
    ytick={-100, -80, -60, -40, -20, 0, 20, 40, 60},
    yticklabel style={anchor=near yticklabel},
    yticklabel={$\pgfmathprintnumber{\tick}\%$},
    ylabel style = {align = center},
    ymajorgrids={true},
    date coordinates in=x,
    xtick={ 
    2020-1-1,
    2020-2-1,
    2020-3-1,
    2020-4-1,
    2020-5-1,
    2020-6-1,
    2020-7-1},
    xticklabel style={rotate=45,anchor=near xticklabel},
    xticklabel={\pgfcalendarmonthshortname{\month}},
    date ZERO=2009-08-18,
	legend pos=north east,
	legend style={at={(.35,0)},anchor=south east,font=\scriptsize},
	enlarge x limits=false,
    ]
    \addplot [blue, mark=square*] table [col sep=comma, x=date, y=bottom]
    {data/NashvilleCensusLinePlot.csv}; \addlegendentry{Low-income},
    \addplot [red, mark=square*] table [col sep=comma, x=date, y=top]
    {data/NashvilleCensusLinePlot.csv}; \addlegendentry{High-income},
    \addplot +[black, dashed, mark=none] coordinates {(2020-3-12, -100) (2020-3-12, 40)};
    \addplot+[black, mark=x, mark options={scale=0}, ultra thick] coordinates {(2020-3-12, -40)} node[pin=170:{Lockdown}]{};
    \addplot +[black, dashed, mark=none] coordinates {(2020-5-11, -100) (2020-5-11, 40)};
    \addplot+[black, mark=x, mark options={scale=0}, ultra thick] coordinates {(2020-5-11, -40)} node[pin=170:{Phase 1}]{};
    \addplot +[black, dashed, mark=none] coordinates {(2020-5-25, -100) (2020-5-25, 40)};
    \addplot+[black, mark=x, mark options={scale=0}, ultra thick] coordinates {(2020-5-25, -20)} node[pin=170:{Phase 2}]{};
    \addplot +[black, dashed, mark=none] coordinates {(2020-6-22, -100) (2020-6-22, 40)};
    \addplot+[black, mark=x, mark options={scale=0}, ultra thick] coordinates {(2020-6-22, -10)} node[pin=95:{Phase 3}]{};
    \addplot+[black, mark=none] coordinates {(2020-1-1, 0) (2020-7-1, 0)};
\end{axis}
\end{tikzpicture}
    \caption{Change in ridership compared to 2019 baseline for the 10\% high income and 10\% lowest income census tracts in Nashville measured by median household income.}
    \label{fig:NashvilleCensus}
\end{figure}

\else
\begin{figure}[t]
    \centering
\pgfplotsset{scaled y ticks=false}
\begin{tikzpicture}
\begin{axis}[
    width=.8\columnwidth,
    height=.35\columnwidth,
    font=\footnotesize,
    ymin = -100, ymax = 40,
    ytick={-100, -80, -60, -40, -20, 0, 20, 40},
    yticklabel style={anchor=near yticklabel},
    yticklabel={$\pgfmathprintnumber{\tick}\%$},
    ylabel={Change in Ridership},
    ylabel style = {align = center},
    ymajorgrids={true},
    xlabel=Month,
    date coordinates in=x,
    xtick={ 
    2020-1-1,
    2020-2-1,
    2020-3-1,
    2020-4-1,
    2020-5-1,
    2020-6-1,
    2020-7-1},
    xticklabel style={rotate=45,anchor=near xticklabel},
    xticklabel={\pgfcalendarmonthshortname{\month}},
    date ZERO=2009-08-18,
	legend pos=north east,
	legend style={at={(.15,0)},anchor=south,font=\scriptsize},
	enlarge x limits=false,
    ]
    \addplot [blue, mark=square*] table [col sep=comma, x=date, y=bottom]
    {data/NashvilleCensusLinePlot.csv}; \addlegendentry{Low-income Group},
    \addplot [red, mark=square*] table [col sep=comma, x=date, y=top]
    {data/NashvilleCensusLinePlot.csv}; \addlegendentry{High-income Group},
    \addplot +[black, dashed, mark=none] coordinates {(2020-3-12, -100) (2020-3-12, 40)};
    \addplot+[black, mark=x, mark options={scale=0}, ultra thick] coordinates {(2020-3-12, -40)} node[pin=170:{Lockdown}]{};
    \addplot +[black, dashed, mark=none] coordinates {(2020-5-11, -100) (2020-5-11, 40)};
    \addplot+[black, mark=x, mark options={scale=0}, ultra thick] coordinates {(2020-5-11, -40)} node[pin=170:{Phase 1}]{};
    \addplot +[black, dashed, mark=none] coordinates {(2020-5-25, -100) (2020-5-25, 40)};
    \addplot+[black, mark=x, mark options={scale=0}, ultra thick] coordinates {(2020-5-25, -20)} node[pin=170:{Phase 2}]{};
    \addplot +[black, dashed, mark=none] coordinates {(2020-6-22, -100) (2020-6-22, 40)};
    \addplot+[black, mark=x, mark options={scale=0}, ultra thick] coordinates {(2020-6-22, -10)} node[pin=95:{Phase 3}]{};
    \addplot+[black, mark=none] coordinates {(2020-1-1, 0) (2020-7-1, 0)};
\end{axis}
\end{tikzpicture}
    \caption{Change in ridership compared to 2019 baseline for the 10\% high income and 10\% lowest income census tracts in Nashville measured by median household income.}
    \label{fig:NashvilleCensus}
\end{figure}
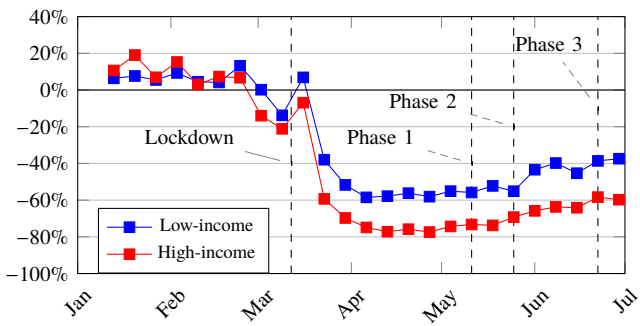
\fi

As we have seen, ridership varies both temporally and spatially in both cities. In this section we investigate the correlation between decreases in ridership and socio-economic factors. As a proxy for wealth we use median household income per census tract. \cref{fig:NashvilleCensus} shows change in weekly ridership for 2020 compared to baseline ridership in 2019 for the 10\% highest income and 10\% lowest income census tracts in Nashville. We see a significantly greater decrease in ridership for the high income group which saw a 77\% decrease in ridership in the week of April 27th. Meanwhile the low-income group saw only a 58\% decrease in the week of April 27th, also a low. The trend lines follow a similar trajectory for both groups, no significant time shift was found. Additionally during post-COVID operations both groups saw similar upward trends in ridership. 

\cref{table:correlation} provides Pearson correlation statistics for various income metrics and racial demographics for ridership drop post-COVID. The moderate positive correlation between median income and housing value reiterates the effects of socio-economic status on ridership post-COVID. In \cref{table:correlation} a positive correlation refers to the case in which as the metric increases the more severe the drop in ridership. The highest positive correlation with drop in ridership was with median housing value, i.e. census tracts with high median housing costs had the greatest reduction in ridership from 2019 baseline. There was also a moderate negative correlation between the percent of the population that was Hispanic and change in ridership. There was no correlation between the percentage of the census tract that was African American or White. 

There are two likely reasons that low-income areas see less of a decrease in public transit usage. First, low-income families are less likely to own a car \cite{klein2017car}. Second, workers in grocery stores, sanitation and cleaning, and logistics are often labeled ``essential'' workers and still required travelling to their place of work. Additionally, as local resources and jobs are often limited in low-income areas \cite{blumenberg2004beyond} travel by public transit is a necessary component of life for these groups, regardless of COVID-19. 

\begin{table}[]
\centering
\caption{Pearson Correlation values for change in ridership after COVID-19 in Nashville Tennessee. A positive correlation refers to as the metric increases, the more severe the drop in ridership post-COVID.}
\begin{tabular}{|l|l|}
\hline
\multicolumn{1}{|c|}{Metric} & \multicolumn{1}{c|}{\begin{tabular}[c]{@{}c@{}}Pearson \\ Correlation\end{tabular}} \\ \hline
Median Income                & 0.21                                                                                \\
Median Housing Value         & 0.35                                                                                \\
Median Rent                  & 0.15                                                                                \\
\% White                     & 0.01                                                                                \\
\% African American          & -0.02                                                                               \\
\% Hispanic                  & -0.19                                                                               \\ \hline
\end{tabular}
\label{table:correlation}
\end{table}



\section{Threats to Validity}\label{sec:threats}

One limitation of this work is that it is focused on two cities, both in Tennessee. Government restrictions vary greatly throughout the United States not only at the state level but at the city level. Even in this study Nashville Metro, the local government of Nashville and Davidson County, systematically enforced restrictions that different from the Tennessee state restrictions under which Chattanooga was regulated. While Nashville has followed an outlined four stage opening plan these stages many have different restrictions compared to other cities and states. Additionally while Nashville had recently moved to a more open stage three in late June it reverted back to stage two by July 4, 2020. This careful approach is necessary to ensure a safe reopening of cities, however from an analysis point of view mixed messaging presents challenges in correlating patterns with ridership demand. However, the findings in this work align with preliminary findings across the United States and the world \cite{mckinsey}.

Secondly, public transit includes confining passengers to an enclosed space whether social distancing is implemented or not. To date, there is no known mass transmission of COVID-19 in Nashville or Chattanooga that originated on public transit. A well publicized case such as this would most certainly have a negative impact on ridership. Historically mass transit can be a source of influenza and coronavirus transmission \cite{browne2016roles} however preliminary findings related to COVID-19 indicate that fears of public transit may be exaggerated \cite{atlantic}. Regardless it is imperative that transit agencies monitor social distancing and put in place adequate sanitation safeguards. 
\section{Conclusion}
\label{sec:conclusion}

In this work we presented a data-driven analysis of the impact of COVID-19 on ridership in Nashville and Chattanooga, TN. We found that ridership dropped by 66.9\% at its peak in Nashville and by July 1, 2020 had stabilized at a 48.4\% reduction compared to 2019 baselines. Chattanooga ridership dropped by 65.1\% at its lowest point and stabilized at 42.8\% off its corresponding 2019~baseline. 

We also showed that the most significant temporal factor in ridership decline occurred during morning and evening commute times. By combining spatial distributions of ridership decline with census level economic data we find that high-income areas of Nashville saw a decreased ridership of more than 19\% compared to low-income areas (77\% for high-income vs 58\% for low-income at their respective lows). A moderate correlation between median income levels and housing value and drop in ridership. 

Future work includes developing low cost image processing methods for ensuring social distancing on public transit. We also plan on using the analysis in this work to set the ground for agent-based simulation and modeling to predict ridership behavior as the COVID-19 pandemic continues to unfold. 

\section*{Acknowledgment}
This work was supported in part by National Science Foundation through award numbers 2029950 and 2029952.  Any opinions, findings, and conclusions or recommendations expressed in this material are those of the author(s) and do not necessarily reflect the views of the National Science~Foundation.

\section{Author contributions}

M. Wilbur and A. Ayman provided technical guidance and management for the research, helped run data analyses, and helped write the manuscript. A. Ouyang, helped with data processing, ran analyses and helped write the manuscript. V. Poon helped with data processing and ran analyses. R. Kabir ran analyses and A. Vadali helped with literature review. P. Pugliese and D. Freudberg helped with data collection and provided technical guidance. A. Laszka and A. Dubey supervised the research and assisted with the manuscript writing.

\bibliographystyle{unsrt}
\bibliography{main}

\begin{thebibliography}{10}

\bibitem{whopandemicannouncement}
{World Health Organization}.
\newblock Who pandemic announcement.
\newblock \url{https://www.who.int/news-room/detail/29-06-2020-covidtimeline}.

\bibitem{a2c2smartwhitepaper2}
Jingqin Gao, Jingxing Wang, Zilin Bian, Suzana~Duran Bernardes, Yanyan Chen,
  Abhinav Bhattacharyya, Siva Soorya~Muruga Thambiran, Kaan Ozbay, Shri Iyer,
  and Xuegang~Jeff Ban.
\newblock The effects of the {COVID-19} pandemic on transportation systems in
  new york city and seattle, usa.
\newblock May 2020.

\bibitem{a12brown2017car}
Anne~E Brown.
\newblock Car-less or car-free? socioeconomic and mobility differences among
  zero-car households.
\newblock {\em Transport Policy}, 60:152--159, 2017.

\bibitem{browne2016roles}
Annie Browne, Sacha St-Onge~Ahmad, Charles~R Beck, and Jonathan~S
  Nguyen-Van-Tam.
\newblock The roles of transportation and transportation hubs in the
  propagation of influenza and coronaviruses: a systematic review.
\newblock {\em Journal of travel medicine}, 23(1):tav002, 2016.

\bibitem{andrews2013modeling}
Jason~R Andrews, Carl Morrow, and Robin Wood.
\newblock Modeling the role of public transportation in sustaining tuberculosis
  transmission in south africa.
\newblock {\em American journal of epidemiology}, 177(6):556--561, 2013.

\bibitem{zhao2013transportation}
Fang Zhao, Thomas Gustafson, et~al.
\newblock Transportation needs of disadvantaged populations: where, when, and
  how?
\newblock Technical report, United States. Federal Transit Administration,
  2013.

\bibitem{bota2017modeling}
Andras Bota, L~Gardner, and Alireza Khani.
\newblock Modeling the spread of infection in public transit networks: A
  decision-support tool for outbreak planning and control.
\newblock In {\em Transportation research board 96th annual meeting}, 2017.

\bibitem{chinazzi2020effect}
Matteo Chinazzi, Jessica~T Davis, Marco Ajelli, Corrado Gioannini, Maria
  Litvinova, Stefano Merler, Ana~Pastore y~Piontti, Kunpeng Mu, Luca Rossi,
  Kaiyuan Sun, et~al.
\newblock The effect of travel restrictions on the spread of the 2019 novel
  coronavirus (covid-19) outbreak.
\newblock {\em Science}, 368(6489):395--400, 2020.

\bibitem{atlantic}
{The Atlantic}.
\newblock Fear of public transit got ahead of the evidence.
\newblock
  \url{https://www.theatlantic.com/ideas/archive/2020/06/fear-transit-bad-cities/612979/},
  June 2020.

\bibitem{a1c2smartwhitepaper1}
Jingqin Gao, Suzana~Duran Bernardes, and Zilin Bian.
\newblock Initial impacts of {COVID-19} on transportation systems: A case study
  of the {U.S.} epicenter, the {New} {York} {Metropolitan} {Area}.
\newblock April 2020.

\bibitem{a3c2smartwhitepaper3}
Ding Wang, Fan Zuo, Jingqin Gao, Yueshuai He, Zilin Bian, Suzana Duran,
  Bernardes, Chaekuk Na, Jingxing Wang, John Petinos, Kaan Ozbay, Joseph~Y.J.
  Chow, Shri Iyer, Hani Nassif, and Xuegang~Jeff Ban.
\newblock Agent-based simulation model and deep learning techniques to evaluate
  and predict transportation trends around {COVID-19}.
\newblock June 2020.

\bibitem{a4c2smartwhitepaper1}
Suzana~Duran Bernardes, Zilin Bian, Siva Sooryaa~Muruga Thambiran, Jingqin,
  Gao, Chaekuk Na, Fan Zuo, Nick Hudanich, Abhinav Bhattacharyya, Kaan Ozbay,
  Shri Iyer, Joseph~Y.J. Chow, and Hani Nassif.
\newblock {NYC} recovery at a glance - the rise of buses and micromobility.
\newblock July 2020.

\bibitem{aloi2020effects}
Alfredo Aloi, Borja Alonso, Juan Benavente, Rub{\'e}n Cordera, Eneko
  Ech{\'a}niz, Felipe Gonz{\'a}lez, Claudio Ladisa, Raquel Lezama-Romanelli,
  {\'A}lvaro L{\'o}pez-Parra, Vittorio Mazzei, et~al.
\newblock Effects of the covid-19 lockdown on urban mobility: Empirical
  evidence from the city of santander (spain).
\newblock {\em Sustainability}, 12(9):3870, 2020.

\bibitem{a5hu2020impacts}
Yue Hu, William Barbour, Samitha Samaranayake, and Dan Work.
\newblock Impacts of {COVID-19} mode shift on road traffic.
\newblock {\em arXiv preprint arXiv:2005.01610}, 2020.

\bibitem{teixeira2020link}
Jo{\~a}o~Filipe Teixeira and Miguel Lopes.
\newblock The link between bike sharing and subway use during the covid-19
  pandemic: The case-study of new york's citi bike.
\newblock {\em Transportation Research Interdisciplinary Perspectives},
  6:100166, 2020.

\bibitem{pucher2003socioeconomics}
John Pucher and John~L Renne.
\newblock Socioeconomics of urban travel. evidence from the 2001 nhts.
\newblock 2003.

\bibitem{a10grahn2019socioeconomic}
Rick Grahn, Chris Hendrickson, Zhen~Sean Qian, and H~Scott Matthews.
\newblock Socioeconomic and usage characteristics of public transit riders in
  the united states.
\newblock Technical report, 2019.

\bibitem{a9apta}
{American Public Transportation Association}.
\newblock Who rides public transportation.
\newblock
  \url{https://www.apta.com/wp-content/uploads/Resources/resources/reportsandpublications/Documents/APTA-Who-Rides-Public-Transportation-2017.pdf}.

\bibitem{a11flannelly1989multivariate}
Kevin~J Flannelly and Malcolm~S McLeod.
\newblock A multivariate analysis of socioeconomic and attitudinal factors
  predicting commuters’ mode of travel.
\newblock {\em Bulletin of the psychonomic society}, 27(1):64--66, 1989.

\bibitem{a13mohamed2014understanding}
K~Mohamed, Etienne C{\^o}me, Johanna Baro, and Latifa Oukhellou.
\newblock Understanding passenger patterns in public transit through smart card
  and socioeconomic data.
\newblock {\em UrbComp,(Seattle, WA, USA)}, 2014.

\bibitem{viriciti}
{ViriCiti}.
\newblock Viriciti electric and non-electric fleet management.
\newblock \url{https://viriciti.com/}.

\bibitem{censusbureau}
{United States Government}.
\newblock United states census bureau.
\newblock \url{https://www.census.gov/}.

\bibitem{proximityone}
{ProximityOne}.
\newblock Proximityone.
\newblock \url{http://proximityone.com/}.

\bibitem{covidcountsnewyorktimes}
{The New York Times}.
\newblock Coronavirus in the u.s.: Latest map and case count.
\newblock
  \url{https://www.nytimes.com/interactive/2020/us/coronavirus-us-cases.html}.

\bibitem{tennhealth}
{Tennessee Office of the Governor}.
\newblock Public health orders.
\newblock \url{https://www.tn.gov/governor/covid-19/covid19timeline.html}.

\bibitem{nashmetrohealth}
{Nashville Metro}.
\newblock Public health orders.
\newblock \url{https://www.asafenashville.org/public-health-orders/}.

\bibitem{tornado}
{USA Today}.
\newblock What we know wednesday about victims of the tennessee tornadoes and
  recovery efforts.
\newblock
  \url{https://www.usatoday.com/story/news/nation/2020/03/04/nashville-tornadoes-what-we-know-deaths-damage-path/4950396002/},
  March 2020.

\bibitem{klein2017car}
Nicholas~J Klein and Michael~J Smart.
\newblock Car today, gone tomorrow: The ephemeral car in low-income, immigrant
  and minority families.
\newblock {\em Transportation}, 44(3):495--510, 2017.

\bibitem{blumenberg2004beyond}
Evelyn Blumenberg and Michael Manville.
\newblock Beyond the spatial mismatch: Welfare recipients and transportation
  policy.
\newblock {\em Journal of Planning Literature}, 19(2):182--205, 2004.

\bibitem{mckinsey}
{McKinsey and Company}.
\newblock Reopening cities after covid-19.
\newblock
  \url{https://www.mckinsey.com/industries/travel-logistics-and-transport-infrastructure/our-insights/reopening-cities-after-covid-19},
  June 2020.

\end{thebibliography}

\end{document}